\documentclass[pra,aps,twocolumn,showpacs]{revtex4}
\usepackage{amsmath,graphicx,epsfig}

\begin{document}

\title{Detection of nuclear magnetic resonance with an anisotropic magnetoresistive sensor} 

\author{F. Verpillat}\email{frederic.verpillat@ens-lyon.fr}
\affiliation{Ecole Normale Sup\'erieure de Lyon, 69364 Lyon cedex 07, France}
\author{M.\ P.\ Ledbetter}\email{ledbetter@berkeley.edu}
\author{D.\ Budker}\email{budker@berkeley.edu}
\affiliation{Department of Physics, University of California at
Berkeley, Berkeley, California 94720-7300}
\author{S. Xu}
\author{D. Michalak}
\author{C. Hilty}
\author{L.-S. Bouchard}
\author{S. Antonijevic}
\author{A. Pines}
\affiliation{Department of Chemistry, University of California at
Berkeley, Berkeley, California 94720}

\date{\today}


\begin{abstract}
We report detection of nuclear magnetic resonance (NMR) using an
anisotropic magnetoresistive (AMR) sensor. A ``remote-detection''
arrangement was used, in which protons in flowing water were
pre-polarized in the field of a superconducting NMR magnet,
adiabatically inverted, and subsequently detected with an AMR sensor
situated downstream from the magnet and the adiabatic inverter. AMR
sensing is well suited for NMR detection in microfluidic
``lab-on-a-chip'' applications.
\end{abstract}



\maketitle

The three essential elements of a nuclear-magnetic-resonance (NMR)
or magnetic-resonance-imaging (MRI) experiment, nuclear-spin
polarization, encoding, and detection can be spatially separated,
which is referred to as remote detection of NMR or MRI
\cite{Mou2003}. One important potential advantage of this approach
is that encoding and detection can occur in a near-zero magnetic
field; however, conventional inductive detection has poor
sensitivity at low frequencies, necessitating the use of alternative
techniques for detection. Superconducting quantum-interference
devices (SQUIDs) \cite{Won2002} and alkali-vapor atomic
magnetometers \cite{Xu2006PNAS} have been successfully utilized for
this purpose. In this note, we report the use of another novel
technology -- that of anisotropic magneto-resistive (AMR) sensors
\cite{Tsy2001} -- for a remote-NMR experiment. Although less
sensitive than SQUIDs or atomic magnetometers (including even the
miniature chip-scale atomic magnetometers \cite{Sch2007}), the
all-solid-state AMR sensors do not require cryogenics or vapor-cell
heating, and may be particularly fit to microfluidic applications
because the sensors are small, inexpensive and can be manufactured
as arrays for spatial sensitivity.

\begin{figure}
  \includegraphics[width=3.5 in]{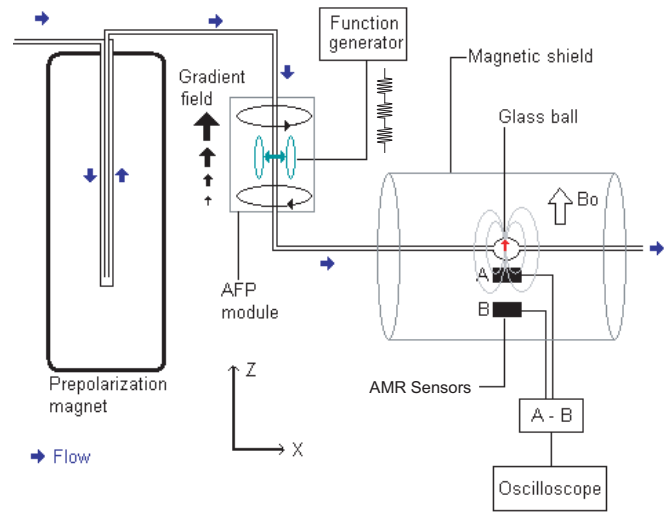}\\
  \caption{Experimental setup. Water is pre-polarized by flowing it through the magnet; the magnetization
  is periodically inverted by passing the liquid through the adiabatic fast passage (AFP)
  module; the magnetization is detected with a gradiometer consisting of two AMR sensors.}\label{setup}
\end{figure}

The experimental setup is shown in Fig. \ref{setup}. Tap water,
pre-polarized by flowing through a Bruker 17 Tesla magnet, flows
through an adiabatic-inversion region, where its polarization is
periodically reversed, after which it proceeds to flow past an AMR
detector. The adiabatic polarization inverter incorporates a set of
coils in anti-Helmoltz configuration to supply a gradient of $\rm
B_z$. A second set of Helmoltz coils is used to apply a $5.5~{\rm
kHz}$ oscillating field in the x direction, resonant with the
protons' Larmor frequency in the center of the inverter. When the
oscillating field is on, as the water flows through the device, its
magnetization is adiabatically reversed. Switching the oscillating
field on and off results in magnetization either parallel or
anti-parallel to the bias field. After the adiabatic inverter, the
water flows into the detection region consisting of a $0.5-{\rm
cm^3}$ glass ball adjacent to a pair of Honeywell HMC1001 AMR
sensors arranged as a gradiometer in order to cancel the common-mode
magnetic-field noise. The active part of the sensor is a thin film
with an area of about ${\rm 1.5~mm\times 1.5~mm}$ packaged in a chip
with dimensions ${\rm 10~mm\times 3.9~mm\times 1.5~mm}$.  The
manufacturer specifications of the HMC1001 sensor give a single-shot
resolution of $40~{\rm \mu G}$ with a read-out rate of 1 kHz,
corresponding to a sensitivity of about $1.8~{\rm \mu G/\sqrt{Hz}}$
assuming white noise.  In our experimental setup, we realized a
sensitivity of about $2.7~{\rm \mu G/\sqrt{Hz}}$ at 20 Hz (per
sensor), however the low-frequency performance was considerably
worse, on the order of ${\rm 40~\mu G/\sqrt{Hz}}$ at 1 Hz,
necessitating long signal averaging. The detection region is housed
inside a single layer of magnetic shielding with open ends. The
water-carrying tube was 1/16'' i.d. and the flow rate was $3.8 ~{\rm
cm^3/s}$, corresponding to an average speed of water of $\approx
2~{\rm m/s}$. The average travel time from the magnet to the
inverter is $\approx 1.5~\rm s$, and it is $\approx 0.5~\rm s$ from
the inverter to the detector.

\begin{figure}
 \includegraphics[width=3.5 in]{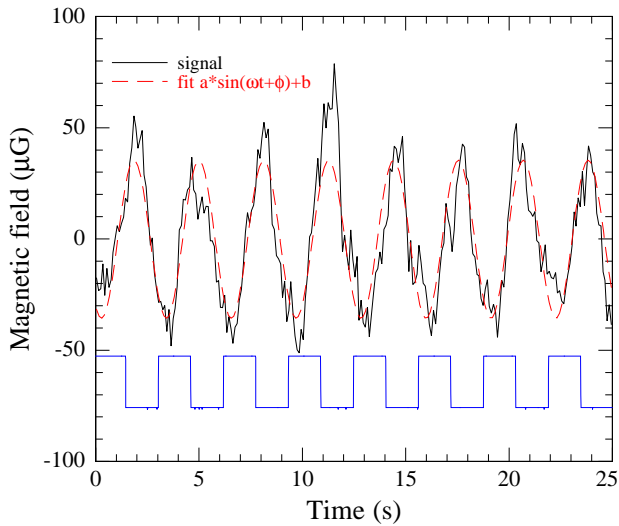}
 \caption{Upper traces -- magnetic field detected with the $0.3-{\rm Hz}$ modulation frequency and a fit to a sinusoidal function.
 Lower trace -- the on-off pattern of the adiabatic inverter. The signal was averaged over 20 min.}\label{signal}
 \end{figure}

Data were recorded on a digital oscilloscope, averaging for about 20
min for modulation frequencies in the range $0.3~{\rm Hz}$ to
$1.7~{\rm Hz}$. There was considerable low-frequency drift in the
signal (due to either ambient field drifting or intrinsic drift in
the magnetometer) and hence we subtracted from the raw data a
slow-varying background approximated by a 3rd-order polynomial. The
resulting signal for a modulation frequency of $0.3~{\rm Hz}$ is
shown in Fig. \ref{signal}. Neglecting fluid mixing in the transfer
tube and detection ball, one would expect that the signal should be
a square wave for low modulation frequencies.  However, considerable
mixing produced signals well approximated (above the cutoff
frequency) by a sinusoid, as indicated by the dashed line in Fig.
\ref{signal}.

\begin{figure}[h]
 \includegraphics[width=3.5 in]{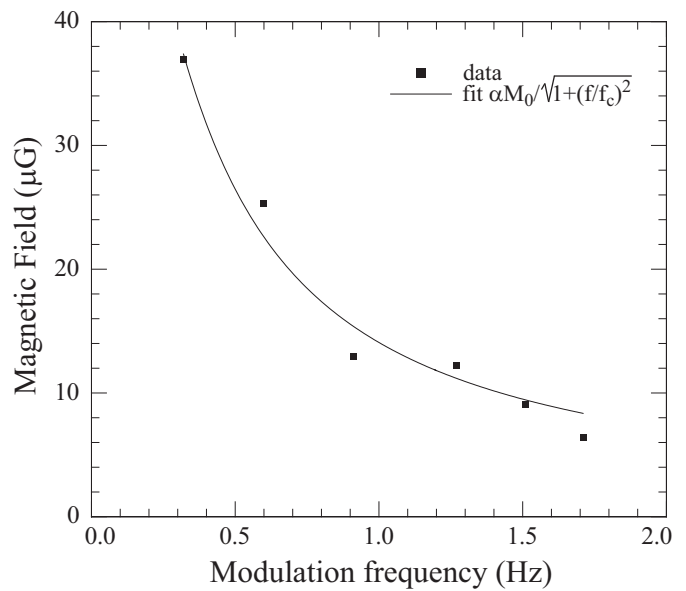}
 \caption{Amplitude of the modulated water signal and a fit to an RC-filter transfer function.}\label{ampl}
\end{figure}

All the data were fit to sinusoidal profiles and the resulting
amplitudes are shown as a function of frequency in Fig.\ \ref{ampl}.
The rapid drop in amplitude is due to mixing of the magnetization as
it propagates from the AFP device through the detection ball,
effectively integrating the magnetization. A simple model for the
spectral response of the system can be obtained in analogy to a
low-pass RC filter where the frequency dependence of the signal is
$S_0=\alpha M_0/\sqrt{1+(f/f_c)^2}$. Here $\alpha$ is a
proportionality constant depending on geometry relating the magnetic
field at the sensor to the magnetization of the sample and $f_c$ is
a cutoff frequency.  Overlaying the data in Fig. \ref{ampl} is a fit
to this model function with $\alpha M_0=67~{\rm \mu G}$ and
$f_c=0.2~{\rm Hz}$.

Significant improvement in sensitivity and bandwidth can be expected
in future work. We suspect that the low-frequency performance of our
AMR sensor was limited by the open-ended magnetic shields used in
the experiment. Optimization of geometry will lead to substantial
gains in both sensitivity to nuclear magnetization (by reducing the
distance from the sample to the magnetometer), as well as improved
bandwidth (by minimizing the volume of the detected water so that
less mixing would occur at high frequencies).  In principle the
detected volume could be a microfluidic channel built into the
sensor package, similar to the construction in Ref. \cite{Pek2004}
where magnetic microparticles were detected.  The higher bandwidth
has the additional benefit of moving the signal above the $1/f$ knee
of the sensor. Higher sensitivity and spatial resolution may also be
achieved by using an array of sensors as in Ref. \cite{Wood2005}.

We have demonstrated, to our knowledge, the first detection of NMR
signals using an anisotropic magneto-resistive sensor. The technique
may be useful for spatial localization (MRI), relaxometry,
diffusometry or spin labeling in chemical analysis \cite{Anw2007}.
With anticipated future advances in AMR sensors, as well as in
related solid-state technologies such as magnetic tunnel junctions,
solid-state chip-scale magnetometers may eventually reach the
picotesla sensitivity level \cite{Fer2006}. Incorporation of
built-in microfluidic channels at the chip level will allow the
construction of dedicated ``lab-on-a-chip'' devices. With these
improvements, room temperature, solid state devices appear to be an
inexpensive and robust alternative for detection of both in-situ and
remote-detection NMR/MRI without cryogenics.

%
%
%
%
%
%


\end{document}